\documentclass[fleqn,usenatbib]{mnras}

\usepackage{newtxtext,newtxmath} 
\usepackage[T1]{fontenc}
\usepackage{ae,aecompl}
\usepackage{graphicx}
\usepackage{amsmath}
\usepackage{amssymb}
\usepackage{epstopdf}

\newcommand{\mpch}{\mbox{Mpc}/\mbox{h}}
\newcommand{\kpch}{\mbox{kpc}/\mbox{h}}
\newcommand{\hmpc}{h^{-1}{\rm Mpc}}

\newcommand{\kms}{{\rm km}/{\rm s}}
\newcommand{\camb}{{\sc CAMB}}

\title[Dipole Distortions in the IGM]{Dipole Distortions in the Intergalactic Medium}
\author[D. Inman, U.-L. Pen and F. Villaescusa-Navarro]{Derek Inman,$^{1}$\thanks{E-mail:
    derek.inman@nyu.edu} Ue-Li Pen$^{2,3,4,5}$\thanks{E-mail: pen@cita.utoronto.ca} and Francisco
  Villaescusa-Navarro$^{6}$\thanks{E-mail: fvillaescusa@flatironinstitute.org}
  \\
  $^{1}$Center for Cosmology and Particle Physics, Department of Physics, New York University, 726 Broadway, New York, NY, 10003, USA\\
  $^{2}$Canadian Institute for Theoretical Astrophysics,
  University of Toronto, 60 St. George St., Toronto, ON M5S 3H8, Canada\\
  $^{3}$Dunlap Institute for Astronomy and
  Astrophysics, University of Toronto, Toronto, ON M5S 3H4, Canada\\
  $^{4}$Canadian Institute for Advanced Research, Program in
  Cosmology and Gravitation\\
  $^{5}$Perimeter Institute for
  Theoretical Physics, Waterloo, ON, N2L 2Y5, Canada\\
  $^{6}$Center for Computational Astrophysics, 162 5th Ave, New York, NY, 10010, USA
}

\date{Accepted XXX. Received YYY; in original form ZZZ}
\pubyear{2018}

\begin{document}
\label{firstpage}
\pagerange{\pageref{firstpage}--\pageref{lastpage}}
\maketitle

\begin{abstract}
  Baryonic feedback can significantly modify the spatial distribution
  of matter on small scales and create a bulk relative velocity
  between the dominant cold dark matter and the hot gas.  We study the
  consequences of such bulk motions using two high resolution
  hydrodynamic simulations, one with no feedback and one with very
  strong feedback.  We find that relative velocities of order
  $100\ \kms$ are produced in the strong feedback simulation whereas
  it is much smaller when there is no feedback.  Such relative motions
  induce dipole distortions to the gas, which we quantify by computing
  the dipole correlation function.  We find halo coordinates and
  velocities are systematically changed in the direction of the
  relative velocity.  Finally, we discuss potential to observe
  the relative velocity via large scale structure, Sunyaev-Zel'dovich
  and line emission measurements.  Given the nonlinear nature of this
  effect, it should next be studied in simulations with different
  feedback implementations/strengths to determine the available model
  space.
\end{abstract}


\section{Introduction}
\label{sec:introduction}

The $\Lambda$CDM model of cosmology is now well established, despite
its two featured ingredients being dark and mysterious.  The observed
accelerated expansion of the Universe implies there must be a dark
energy that behaves at least somewhat like a cosmological constant.
There must also be highly clustered dark matter whose dynamics are
mostly similar to those of a perfectly cold kinetic particle except
perhaps at the smallest scales.  From this perspective, there is then
very little about the macroscopic Universe we {\it don't} understand
since these two components make up $95\%$ of its energy.

Fortunately, the luminous $5\%$ of the Universe, the baryons, provides
a host of their own mysteries.  Observations of galaxies and their
stars tells us that the stellar matter can only account for $\sim10\%$
of the baryonic matter and the location of the remaining $\sim90\%$,
and why it isn't in stars, constitutes the ``missing baryons
problem.''  Galactic feedback processes, which can heat and expel
baryons as a warm-hot intergalactic medium, are thought to resolve
this problem \citep{bib:Tornatore2010}. We believe that feedback comes
mainly in two flavors, arising from different physical processes and
involving different energy scales. On one hand, supernova feedback
involves small scales and is responsible for the suppression of the
stellar mass function for small galaxies. On the other, Active
Galactic Nuclei (AGN) feedback operates on large scales and is
commonly believed to be the mechanism responsible for the stellar mass
function suppression for large galaxies.

Unfortunately, we do not have a complete physical understanding of
these phenomena. In cosmological hydrodynamic simulations the large
range of scales involved do not allow us to simulate these processes
directly and instead subgrid models have been developed to capture
them in a phenomenological, but physically motivated, manner (see the
Horizon \citep{bib:Dubois2014}, Illustris
\citep{bib:Vogelsberger2014b}, IllustrisTNG \citep{bib:IllustrisTNG}
or Eagle \citep{bib:Schaye2015} simulations as relevant examples).
The ``missing baryons'' can then be found in simulations.  As an
example, \citet{bib:Haider2016} found that around one quarter of
baryons end up in halos, just under a half in filaments, and nearly a
third in voids.  These fractions are not replicated without feedback
where more baryons end up in halos.  Additionally, the large numbers
of void baryons was found to be due to feedback.

\begin{figure*}
  \includegraphics[width=\textwidth]{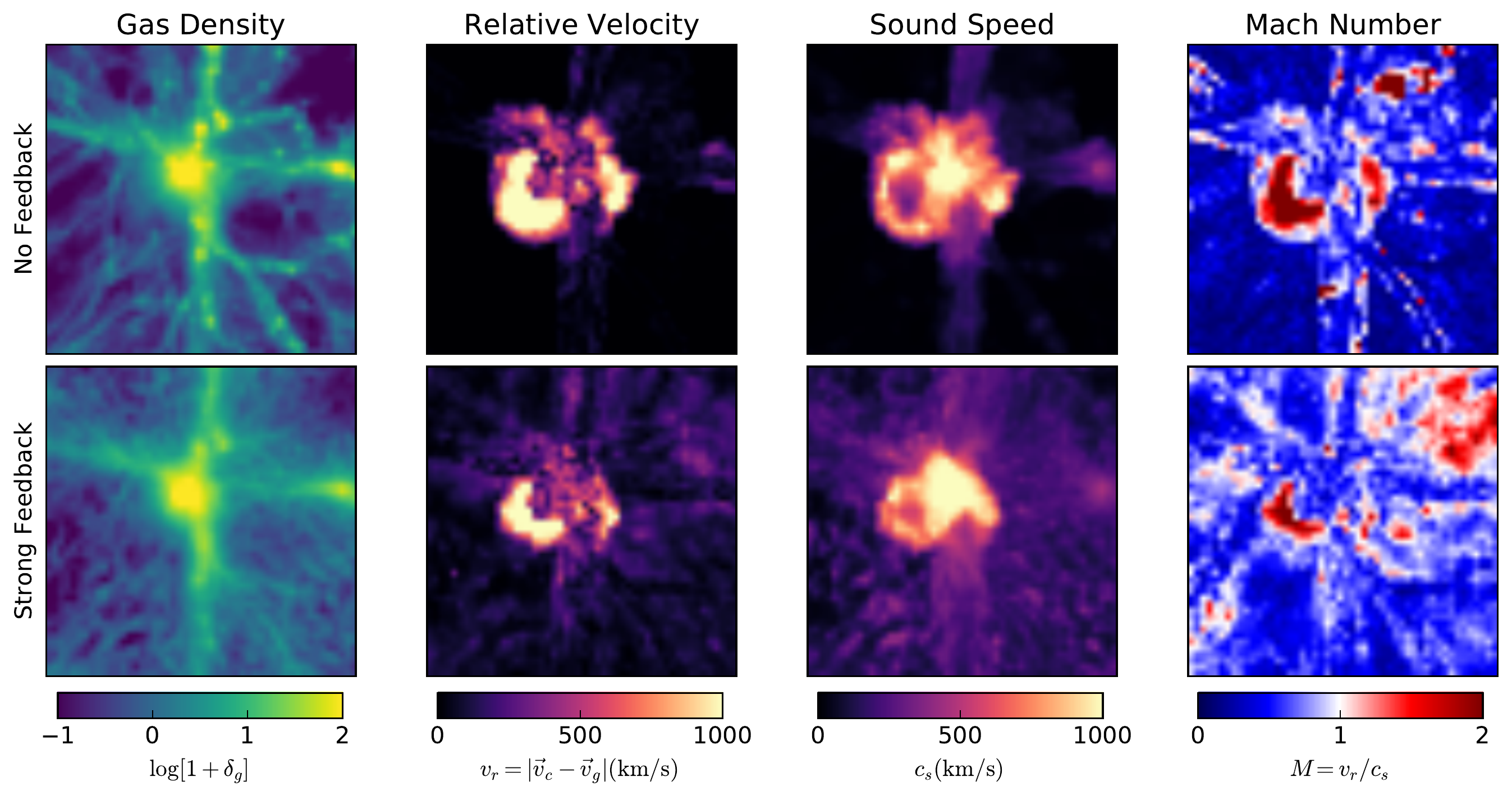}
  \caption{Slices of the gas density, the modulus of the relative
    velocity, sound speed and Mach number.  The slices shown here are
    at redshift $0$ and have lengths of $39$ \mpch{} and depth $4$
    \mpch.  The upper row corresponds to a simulation with no feedback
    whereas the bottom row has $50\times$ enhanced feedback.  When
    feedback is strong, the central halo can expel gas, leading to
    large relative velocities outside the halo.}
  \label{fig:slices}
\end{figure*}

On the other hand, these feedback processes will affect not just the
host galaxy (and halo) itself, but can propagate through the entire
intergalactic medium (IGM) as well \citep{bib:Vogelsberger2014a}.  We
can therefore attempt to learn about feedback through the study of
large scale structure via standard clustering statistics such as the
power spectrum in Fourier space or correlation function in
configuration space \citep{bib:Springel2017}.  An alternative
possibility is to look for direct dynamical consequences of feedback:
energy injection and subsequent heating of the IGM, alongside ejection
of the gas out of galaxies.  These effects introduce a scale below
which there can be a relative velocity between the cold dark matter
(CDM) and the gas.  A halo moving through the gas will then induce a
dipolar density distortion which can then back react and slow down the
CDM.

This paper explores the consequences of the relative velocity between
the baryons and CDM, as a function of feedback energy. We study this
effect by running two simulations with the same initial conditions and
gravitational evolution, but differing strength of feedback.  We find
that the feedback induced relative velocity is largest at redshift
$1$, and use this as our fiducial redshift. We begin by quantifying
the dipole distortion in the IGM by computing the ``dipole correlation
function''.  We then compute the backreaction effect by comparing
individual halos in the two simulations.  Finally, we discuss how this
feedback driven effect can potentially be observed using measurements
of the Sunyaev-Zel'dovich effect, X-ray emission and nonlinear
reconstruction.

This paper is organized as follows. In section \ref{sec:simulations}
we describe the hydrodynamic simulations we have run, and the method
employed to compute the density, velocity and sound speed fields. We
present the main results in section \ref{sec:results} and we draw and
discuss the main conclusions of this work in section
\ref{sec:conclusions}.

\section{Simulations}
\label{sec:simulations}

We run high resolution hydrodynamic simulations using the SPH-TreePM
code {\sc Gadget-III} (see \citet{bib:Springel2001} and
\citet{bib:Springel2005} for descriptions of previous versions of the
code) containing $1024^3$ cold dark matter particles and $1024^3$
baryonic Smooth-Particle Hydrodynamic (SPH) gas particles in a box of
side length $200\ \mpch$.  Our cosmological parameters are consistent
with Planck cosmology: $\Omega_m = 0.3175$,
$\Omega_\Lambda = 1 - \Omega_m = 0.6825$, $n_s=0.9625$, $h=0.6711$ and
$\sigma_8=0.834$ \citep{Planck_2018}. The initial conditions have been
generated at $z=99$ by displacing the positions of CDM and baryons
from two uniform, but offset with respect to one another, grids using
the Zel'dovich approximation. We take into account that the growth
factors/rates of both components are different when computing the
initial displacements and peculiar velocities \citep{Schmidt_2016,
  Valkenburg_2017, Zennaro_2017}. The $z=0$ matter power spectra and
transfer functions are computed using \camb\footnote{http://camb.info}
\citep{bib:Lewis2000}.

Our simulations include radiative cooling by hydrogen and helium and
heating by a uniform UV background. They account for both star
formation and supernova feedback as described in
\citet{bib:Springel2003}. Stellar winds are ejected isotropically from
a star particle.  The parameter we vary between simulations is the
efficiency of the feedback, $\chi$, which describes the fraction of
supernova energy which goes into winds.  Nominally $\chi \leq 1$; on
the other hand, we expect that the kinetic winds underestimate the
energy of real galactic feedback which are also driven by AGN. We
therefore allow $\chi$ to exceed $100\%$ efficiency and use $\chi=0$
and $\chi=50$, which we will refer to as ``no feedback'' and ``strong
feedback'' respectively.  The corresponding wind speeds are given by
$v_w(\chi)=490\sqrt{\chi}\ \kms$.  The wind implementation turns off
the hydrodynamical force of the wind particle for a certain amount of
time or distance traveled as otherwise the gas tends to stay in the
galaxy rather than be ejected.  However, this can significantly affect
galaxy properties \citep{bib:DallaVecchia2008}.  For $\chi=50$, we
tried turning this feature off (so the wind particles always feel a
hydrodynamical force) and found the results are largely similar to the
standard procedure, since for these large velocities the wind almost
immediately re-couples.  We therefore do not expect this to
qualitatively affect our results, although it could affect the
quantitative details. We emphasize that the purpose of this work is
not to implement realistic models of supernova or AGN feedback, but to
study whether astrophysical phenomena that produce winds/jets with
velocities ranging from hundreds to thousand of km/s can induce a
dipole between CDM and the gas.

Of crucial importance to our work is the density, velocity and
hydrodynamic sound speed as a function of position.  We show slices of
these fields in Fig.~\ref{fig:slices}.  The effects of increased
feedback efficiency are clear: the gas density is less clustered and
there are significant bulk flows away from large scale structure.
While these flows are primarily subsonic, there are regions of
supersonic flow as well.  In the next sections we describe how we
compute each field.

\subsection{Density}
\label{ssec:density}
\begin{figure}
  \includegraphics[width=\columnwidth]{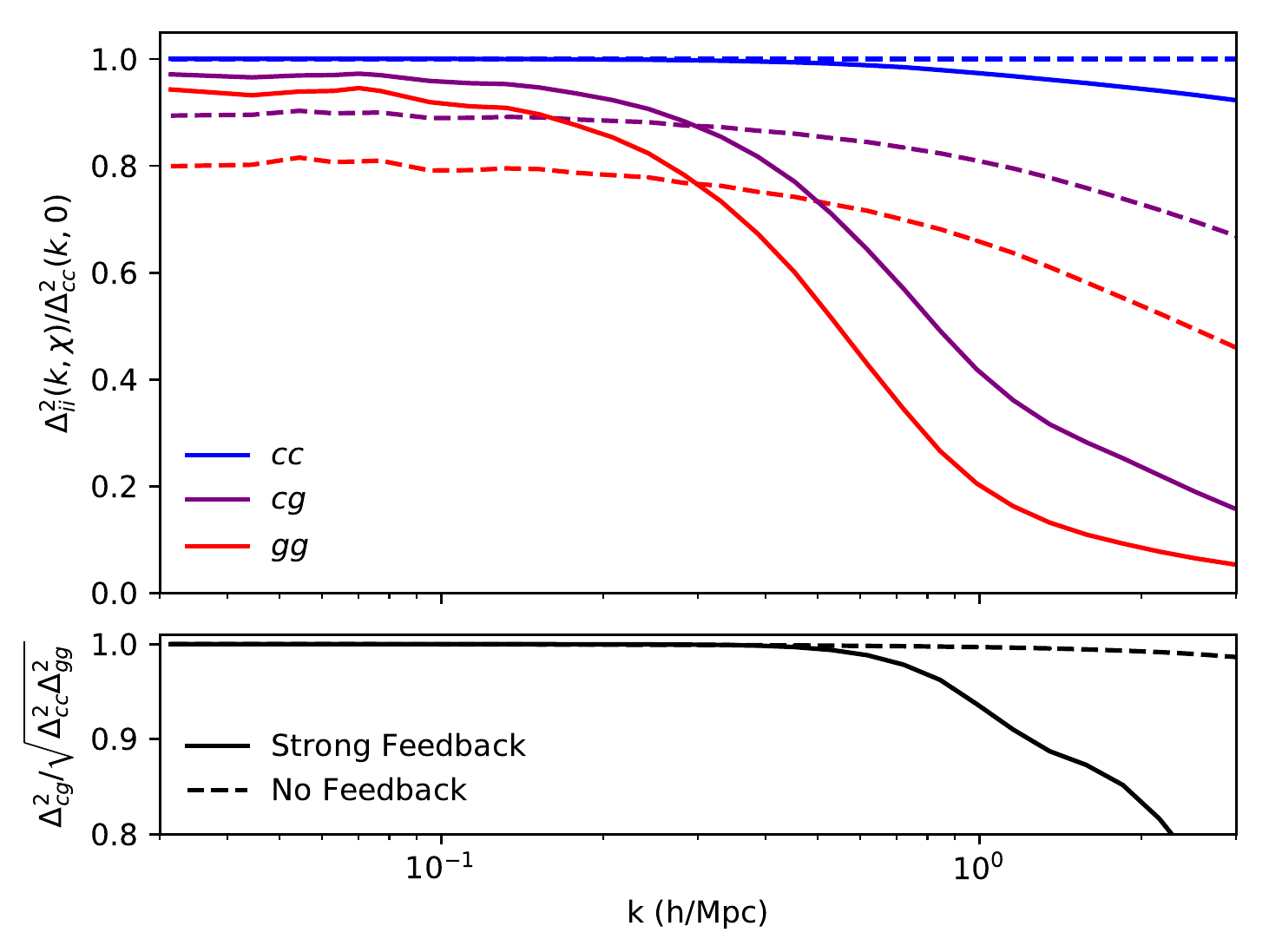}
  \caption{(Top) Ratio of CDM, gas and cross power spectra of the
    simulations with strong feedback to the CDM power spectrum without
    feedback.  The gas power is cutoff on small scales due to its
    hydrodynamical pressure which is absent for CDM.  Strong feedback
    reduces clustering in both the gas and CDM. (Bottom) The
    cross-correlation coefficient between gas and CDM.  Strong
    feedback causes the gas to diverge from the CDM on small scales.}
  \label{fig:denpow}
\end{figure}

We use the standard Cloud-In-Cell (CIC) technique to interpolate CDM
and gas particles to $256^3$ cell cubical grids.  Visually, we see in
Fig.~\ref{fig:slices} that the gas is ``puffier'' than typical CDM
structures and also that feedback further prevents clustering.  To
quantify this more robustly, we compute the dimensionless power
spectra, $\Delta^2_{ij}(k) = \frac{k^3}{2\pi^2}P_{ij}(k)$ where
$P_{ij} = \langle \delta_i(k)\delta_j^*(k) \rangle$ and $i,j=c,g$ for
CDM and gas.  The top panel of Fig.~\ref{fig:denpow} shows the ratio
of power spectra relative to the no feedback CDM power spectrum.  The
gas does not match the large scale CDM as some of the gas particles
are converted to stars (see \citet{bib:Jing2006} and
\citet{bib:Rudd2008} for discussions).  Feedback significantly reduces
the number of stars in clustered regions, so the large scale value of
the strong feedback gas power is closer to that of the CDM.  On small
scales, we see a characteristic cutoff in the gas power spectrum due
to the thermal support of the gas.  Increased feedback heats the gas
causing this cutoff to occur at larger scales than without winds.
Finally, we also note that the CDM power spectrum is also slightly
suppressed by strong feedback.  The bottom panel shows the
cross-correlation coefficient between gas and CDM.  As expected, the
effect of feedback is to decohere the gas from the CDM.

\subsection{Velocity Fields}
\label{ssec:velocity}
Computing velocity fields from a set of discrete particles is a known
numerical challenge.  The simulation momentum field can always be
robustly estimated via the CIC interpolation technique, however,
defining the velocity field as the ratio of the momentum to density
fields is not well defined in cells with no particles.  To our
knowledge, the best way to compute velocity fields is the phase space
projection method performed in \citet{bib:Hahn2015} which was shown to
perform better on small scales than other common schemes such as
Delauney tessellation.  The downside of utilizing the phase space
structure of CDM is that it requires the ability to trace particles
back to their initial conditions which requires extra memory and
computational time.  Another robust option is the
``N-Nearest-Particles'' method (NNP) which defines the velocity of a
grid cell as the average of the N nearest particles to the grid cell.
The 1NP method has been studied theoretically and numerically in
\citet{bib:Zhang2015,bib:Zheng2015} and higher N has been used in
\citet{bib:Inman2015} to average over the random thermal velocities
present in neutrino particles.  While this technique is helpful in
estimating velocities in low density regions, it has the downside that
it does not use all particle information in high density regions.

Fortunately for us, we are using a rather coarse grid and the particle
number density (64 particles per cubic grid cell) is very high.  We
can therefore get away with the use of CIC interpolation, setting any
empty cells to zero. On lower resolution simulations with $512^3$
particles, we have checked that CIC gives comparable results to
8NP. We do not correct the velocity modes amplitude in Fourier space
to remove the convolution with the CIC kernel function, as its
functional form is not trivial when computing the velocity as the
ratio of the momentum over the density. In order to verify that the
CIC mass assignment scheme does not introduce numerical artifacts, we
have computed the velocity field of the gas using the SPH radii of the
particles (see e.g. \cite{Paco_2014}). The velocity power spectrum of
the gas was almost identical with both methods, demonstrating that our
results are robust against these choices.

\begin{figure}
  \includegraphics[width=\columnwidth]{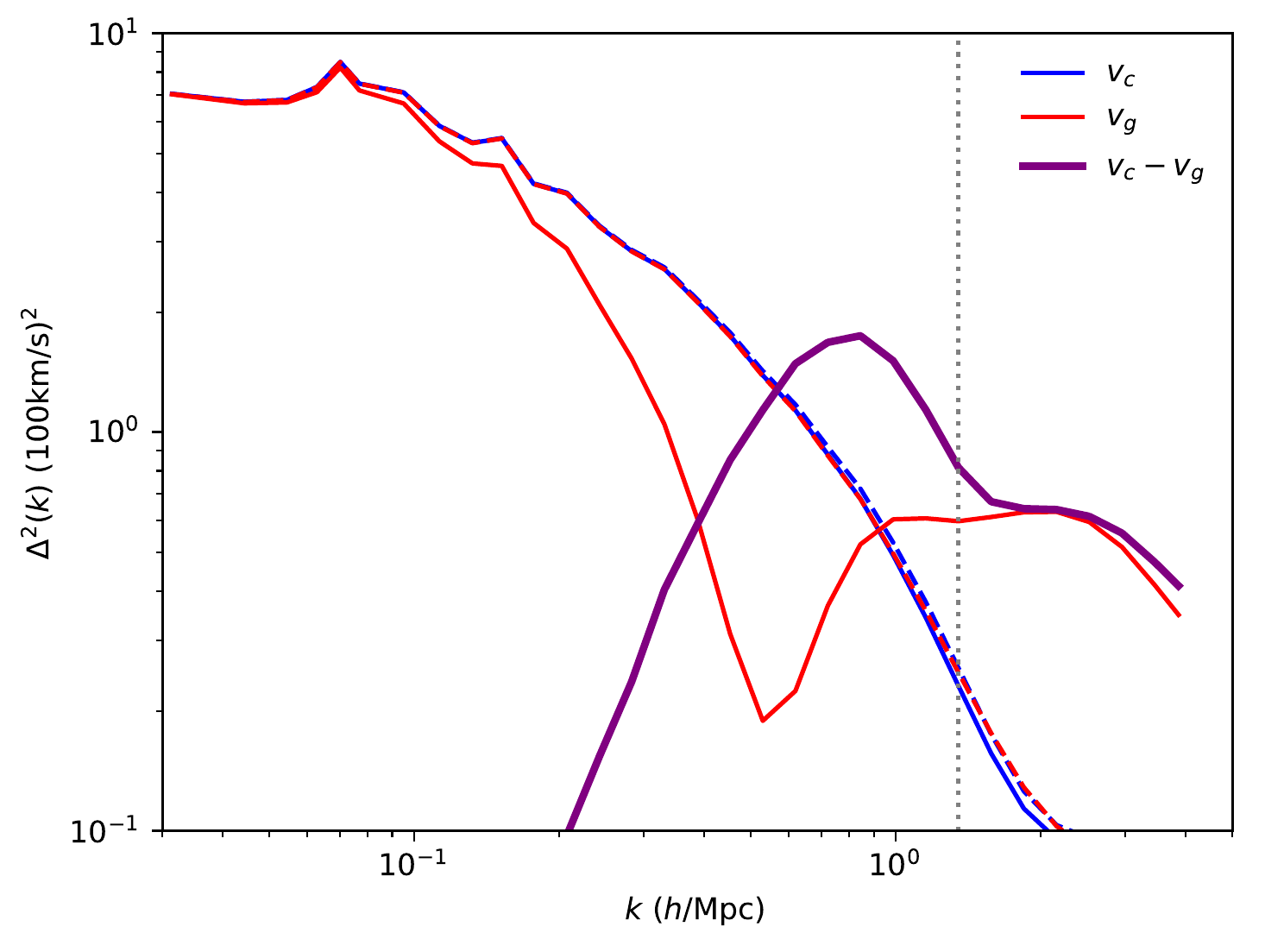}
  \caption{CDM (blue), gas (red), and relative (purple) velocity power
    spectra.  Dashed curves have no feedback whereas solid curves have
    strong feedback.  The suppression in the gas velocity power
    compared to CDM one on nonlinear scales leads to bulk relative
    motions.  We suspect that features on scales smaller than the
    dotted vertical line (i.e.~$k\gtrsim1.35~h^{-1}$Mpc) are
    artificial.}
  \label{fig:relvelpow}
\end{figure}

We show the CDM, gas, and relative velocity power spectra in
Fig.~\ref{fig:relvelpow}, where the velocity power spectrum is defined
as $\Delta^2_{v_x}+\Delta^2_{v_y}+\Delta^2_{v_z}$.  We see that the
CDM velocity power spectra are more or less the same in the two
simulations whereas the gas power spectrum is significantly suppressed
via feedback.  This can be seen directly in the second column of
Fig.~\ref{fig:slices} where relative motions extend to much larger
scales when feedback is turned on.  This difference leads to the large
relative velocity power spectrum with feedback.  We note that there is
a bump in all the velocity power spectra at high $k\gtrsim 1.35 \hmpc$
(most notably in the gas spectra with feedback). This does not seem to
occur in the power spectra computed by \cite{bib:Hahn2015} so we
suspect it may be artificial and likely due to Poisson fluctuations.

\begin{figure}
  \includegraphics[width=\columnwidth]{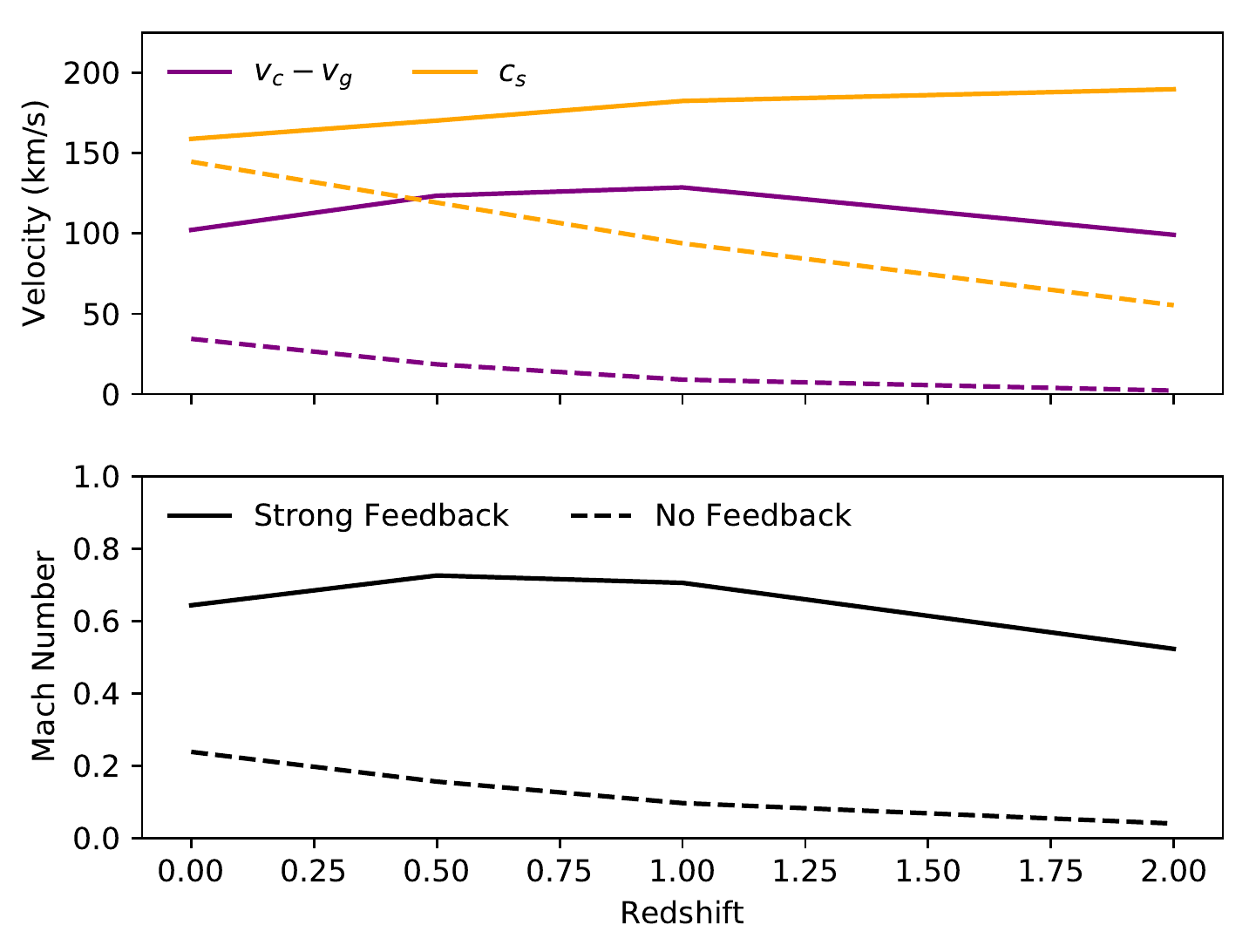}
  \caption{(Top) The variance of the relative velocity (purple) and
    gas sound speed (orange) as a function of redshift.  Strong
    feedback (solid) increases both quantities significantly relative
    to no feedback (dashed).  (Bottom) The Mach number, defined as the
    ratio of the relative velocity to the sound speed.}
  \label{fig:vcz}
\end{figure}

To give a feel for the typical numbers involved, we compute the
variance of the velocity fields,
$\sigma_{i}^2 = \int \Delta^2_i(k) d\ln(k)$.  In order to not be
sensitive to the artificial upturn in the velocity power, we only
integrate to a maximum wavenumber of $1.35~\hmpc$.  We show our
results as a function of redshift in Fig.~\ref{fig:vcz} as the purple
curves.  The feedback induced relative velocity peaks around redshift
$1$ and is always significantly enhanced relative to the simulation
without feedback.  We see that their is a decrement in gas velocity
leading to an increase in relative motions due to feedback.

\subsection{Sound Speed}
\label{ssec:csp}
Our next goal is to estimate the typical size of the gas sound speed
to determine whether relative motions are supersonic or subsonic.
Each SPH particle has an associated internal energy parameter, $u$.
We interpolate the energy to a grid using CIC interpolation,
consistent with our previous calculations.  The sound speed is then
calculated in each cell via $c_s = \sqrt{\gamma(\gamma-1)u}$ where
$\gamma=5/3$ is the ratio of specific heats for a monatomic gas.
Slices of sound speed are shown in the third column of
Fig.~\ref{fig:slices}.  It clearly shares similar features to the
relative velocity, but is not quite the same, leading to flows with
different Mach numbers.  We compute the variance of the internal
energy and take $\sigma_{c_s}=\sqrt{\gamma(\gamma-1)\sigma_u}$.  This
result is shown in the top panel of Fig.~\ref{fig:vcz} as solid
(strong feedback) and dashed (no feedback) orange curves.  The sound
speed is always greater than the relative velocity.  Thus, we expect
on average there to be predominantly subsonic flows with typical Mach
numbers shown in the bottom panel of Fig.~\ref{fig:vcz}.  Nonetheless,
we do still see regions of supersonic flows in the rightmost panels of
Fig.~\ref{fig:slices}.

\section{Consequences for Large Scale Structure}
\label{sec:results}

\subsection{Dipole Correlation Function}
A perturber, such as a CDM halo, moving through a gas will cause wakes
to form downstream from its motion.  These distortions are asymmetric
and depend on the direction of the relative motion.  They are
therefore not captured by the standard two point correlation function
which only contains information about the matter distribution in
spherical shells.  Instead, we can use a three point function, the
``dipole correlation function'' which quantifies how likely you are to
find a subsequent perturbation in a particular direction.

\begin{figure}
  \includegraphics[width=\columnwidth]{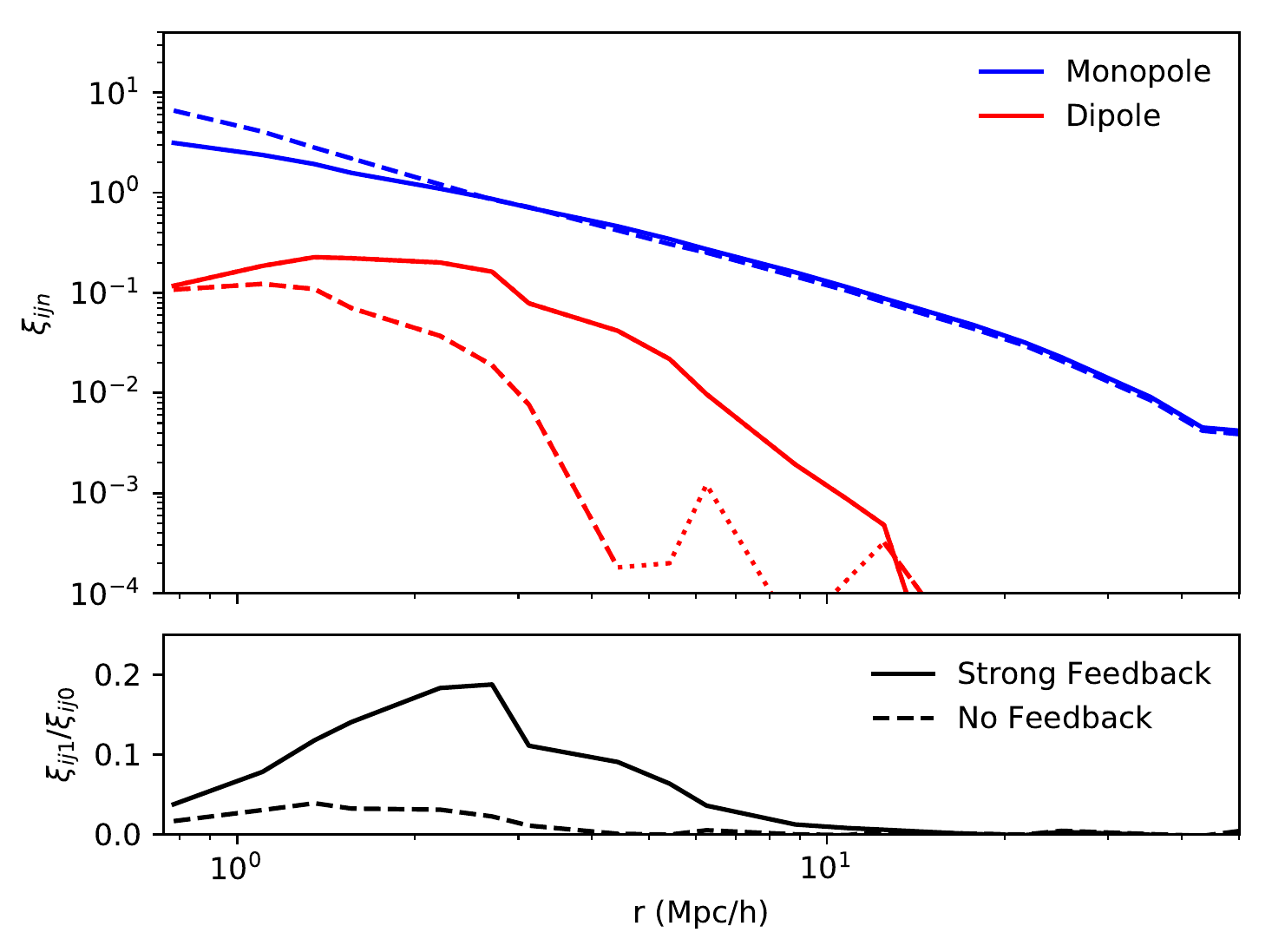}
  \caption{(Top) The monopole (blue) and dipole (red) correlation
    functions between cold dark matter and gas computed from
    simulations.  Solid curves feature strong feedback whereas dashed
    curves have no feedback.  The dotted red curve is the absolute
    value of the dipole with no feedback.  The monopole is suppressed
    on small scales, whereas the dipole is enhanced on larger ones.
    (Bottom) The ratio between monopole and dipole for strong (solid)
    and no (dashed) feedback.  Feedback enhances the polarizability of
    the IGM on scales of a few $h^{-1}{\rm Mpc}$.}
  \label{fig:xi}
\end{figure}

We compute the monopole and dipole correlation functions as
$\xi_{ij0}(r) = \langle
\delta_i(\vec{x})\delta_j(\vec{x}+\vec{r})\rangle$
and
$\xi_{ij1}(r) = \langle
\delta_i(\vec{x})\delta_j(\vec{x}+\vec{r})\hat{v}_{\rm rel}^{\rm
  avg}(\vec{x},\vec{x}+\vec{r})\cdot\hat{r} \rangle$,
where $\hat{v}^{\rm avg}$ is the (unit) average relative velocity at
$\vec{x}$ and $\vec{x}+\vec{r}$,
i.e.~$\vec{v}^{\rm avg}(\vec{x},\vec{x}+\vec{r}) =
(1/2)(\vec{v}(\vec{x})+\vec{v}(\vec{x}+\vec{r}))$.
The algorithm describing this computation is given in
\citet{bib:Inman2017b}.  We show the CDM-gas monopole and dipole
correlation functions in the top panel of Fig.~\ref{fig:xi}.  The
monopole is essentially unchanged by feedback on large scales, but is
slightly suppressed on the small ones.  The dipole is the opposite:
feedback enhances the dipole by a significant amount for $r\sim10$
\mpch.  We also see that strong feedback causes the dipole to extend
to much larger scales.  This makes sense: as we have seen, the gas is
much less clustered with strong feedback.  In the bottom panel of
Fig.~\ref{fig:xi} we show the polarizability, the ratio of the dipole
to the monopole.  We find that it peaks on scales around a few \mpch.
We note that there is significant redshift evolution of the dipole
correlation function.  For instance, at redshift $2$, the strong
feedback $\xi_{ij1}$ is always an order of magnitude larger than the
no feedback one, regardless of scale.  On the other hand, at redshift
$0$ they become much more comparable.  This could be due to the small
amount of relative motion in the no feedback simulation that is
growing with scalefactor, in addition to increasing contamination from
nonlinear gravitational evolution of the density fields.

\subsection{Halo Dynamics}
\label{ssec:friction}
The large bulk motions induced by feedback should change the
coordinates and velocities of halos relative to the no feedback
simulation.  By comparing the same halos between simulations, we can
therefore estimate the net displacement and deceleration.  The most
robust way to match halos between simulations is to check that they
contain the same particles, as has been performed in
\cite{bib:Mummery2017}.  We instead opt for a simpler method to match
halos by enforcing that they are sufficiently similar in mass and
position.  Halo $i$ in one simulation is considered matched to halo
$j$ in the other if $| m_i - m_j |/\sqrt{m_im_j} \le 0.5$ and
$|\vec{r}_i -\vec{r}_j| \le\ 0.5~\mpch.$ In the event that several
halos satisfy these criteria, we select the closest one.  We then
repeat this procedure but swapping the simulations.  We only consider
halos that match each other and have masses greater than
$5\times 10^{11} M_\odot$ corresponding to $\sim 1000$ or more CDM
particles.  In total, we match over $80\%$ of halos. In order to
confirm the robustness of our method we have paired a fraction of the
total halos by matching the IDs of their belonging particles. We find
that the two methods produce very similar results.

Given matched halos, we then compute the difference in their positions
and velocities, $\delta \vec{x} = \vec{x}(\chi) - \vec{x}(0)$ and
$\delta \vec{v} = \vec{v}(\chi) - \vec{v}(0)$, where $\chi$ and 0
stand for the simulations with and without feedback, respectively.
Between the simulations, we expect intrinsic differences in these
quantities that are not directly due to the relative velocity external
to the halos (e.g.~the change in star formation, the change in halo
mass and correspondingly different gravitational accelerations).  To
isolate dynamics due to larger scale flows, we need an estimate of the
relative velocity.  Rather than use the grid velocity fields, which
would necessarily introduce anisotropies, we estimate the gas and CDM
velocities in spheres of radius $5\ \mpch$ around each halo.
Specifically, we compute the mean CDM velocity from all CDM particles
and the mean internal energy and gas velocity from all SPH particles
within the sphere.  We then obtain $\vec{v}_r=\vec{v}_c - \vec{v}_g$
for all halos and define
$\delta \vec{v}_r=\vec{v}_r(\chi)-\vec{v}_r(0)$.

\begin{figure}
  \includegraphics[width=\columnwidth]{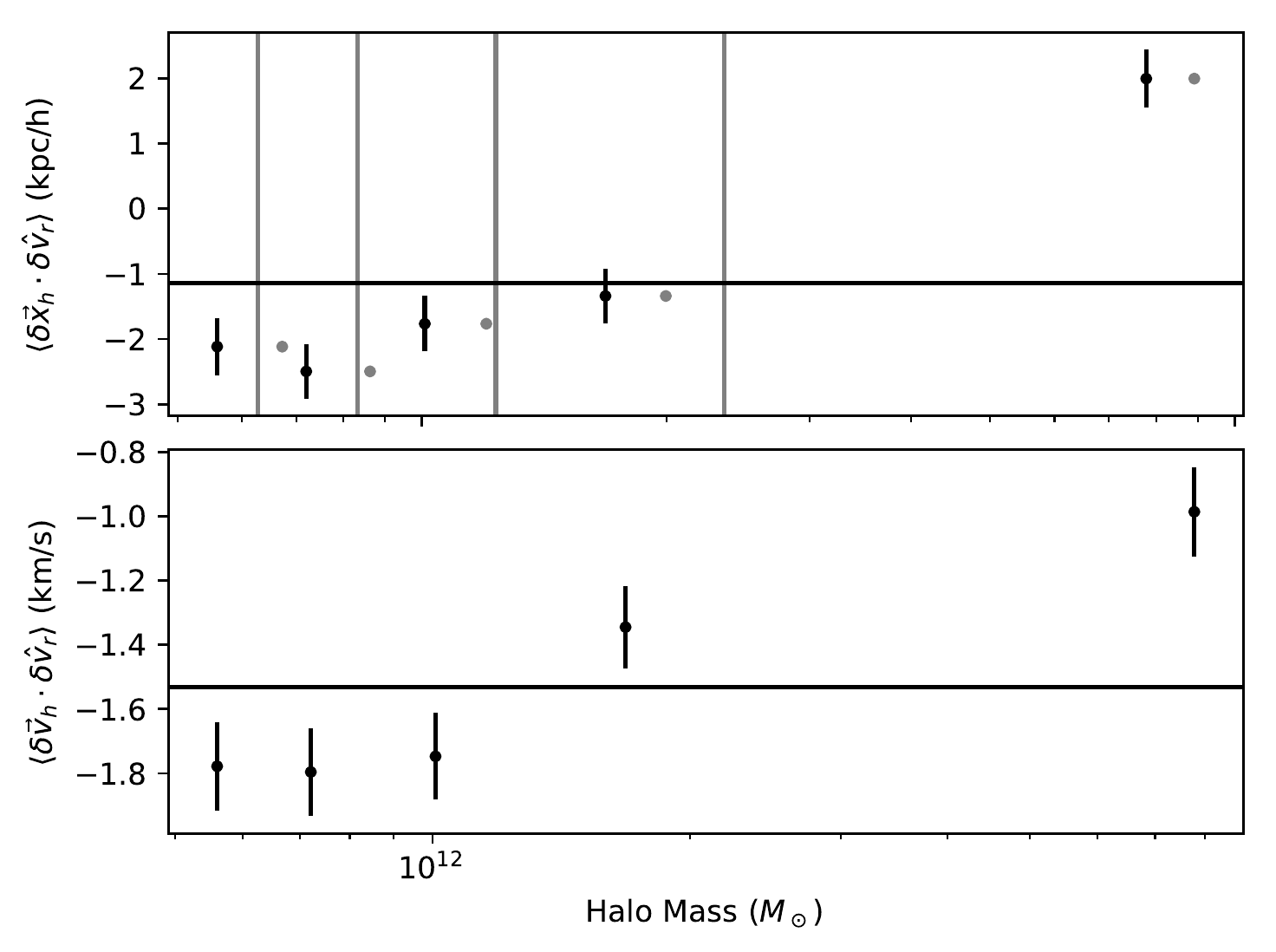}
  \caption{The change in halo positions (top) and velocities (bottom)
    in the direction of the relative velocity as a function of halo
    mass ($M\gtrsim 5\times10^{11}M_\odot$) in the strong feedback
    simulation.  The top panel also shows the mass bins as vertical
    lines and indicates the no feedback halo mass by grey points.
    Horizontal lines show the mean value of all halos
    considered. Potential causes for the acceleration include
    dynamical friction, Ram pressure, and gravitational infall while
    the mass dependence may be affected by halo environment.}
  \label{fig:deltaxv}
\end{figure}

\begin{figure}
  \includegraphics[width=\columnwidth]{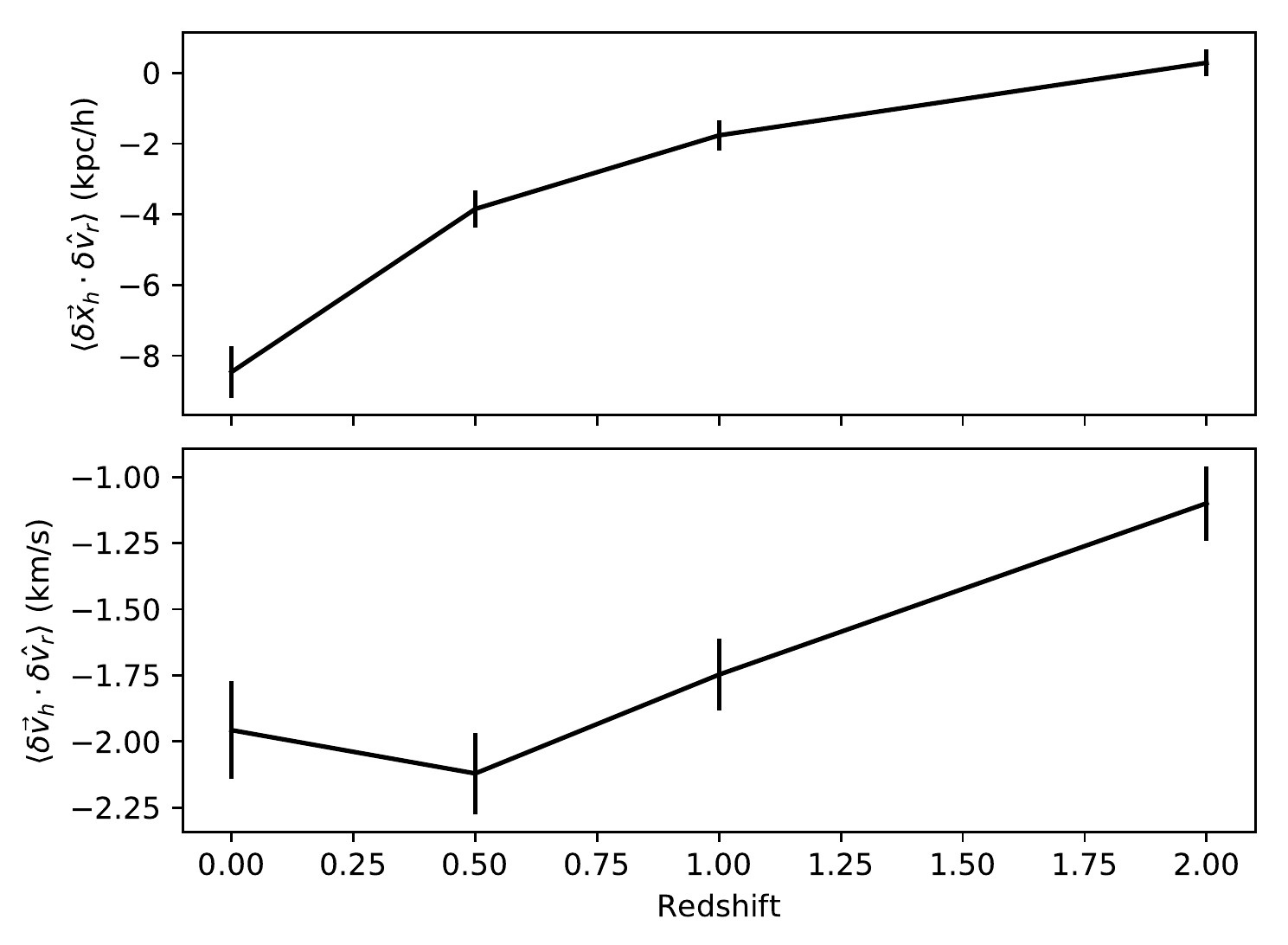}
  \caption{The mean change in halo positions (top) and velocities
    (bottom) in the direction of the relative velocity as a function
    of redshift.  We include only halos with masses around
    $10^{12}M_\odot$ in this computation.}
  \label{fig:xvz}
\end{figure}

We next estimate the change in halo positions and velocities in the
direction of the relative velocity;
$\langle \delta \vec{x}\cdot \delta\hat{v}_r \rangle$ and
$\langle \delta \vec{v} \cdot \delta\hat{v}_r \rangle$, respectively.
We show the results as a function of mass in Fig.~\ref{fig:deltaxv}.
We find that smaller halos tend to be displaced/decelerated further in
the direction of the relative velocity than larger ones.  Indeed we
find that larger halos tend to move in the opposite direction
(i.e.~into the flow rather than with the flow) compared to smaller
ones.  We then select the third bin, corresponding to halo masses
around $\sim10^{12} M_\odot$ and show the results as a function of
redshift in Fig.~\ref{fig:xvz}.  Note that the errors in both figures
are the errors on the mean, not the variance.  This is only a part of
the intrinsic variation in halo coordinates between simulations (
$\sqrt{\langle\delta \vec{x}^2\rangle} \sim \mathcal{O}(100\ \kpch{})
$
and
$\sqrt{\langle\delta \vec{v}^2\rangle}\sim \mathcal{O}(20\ \kms)$).

There are a number of potential mechanisms that can contribute to
these accelerations.  A hydrodynamic explanation is that gas flowing
over the halo can push the gas in the halo via Ram pressure
\citep{bib:Gunn1972}.  Alternatively, it could be entirely
gravitational in nature.  The gas flowing over a halo should be
focused into a downstream wake via dynamical friction, which then
pulls on the halo.  We note that it is theoretically expected that
uniform subsonic flows should not accelerate an isolated point-like
halo \citep{bib:Ostriker1999}.  We discuss this prediction further in
Appendix \ref{sec:exist} and conclude that it is unlikely to hold in
the cosmic web.  Both Ram pressure and dynamical friction will yield
negative values of
$\langle \delta\vec{x}_h\cdot\delta\hat{v}_r \rangle$ and
$\langle \delta\vec{v}_h\cdot\delta\hat{v}_r \rangle$.  On the other
hand, if the gas flows towards halos, but not over them, then there is
just mutual gravitational infall, which will contribute the opposite
sign.  There are as well potential systematic effects.  For instance,
we do not consider the magnitude (and whether it depends on halo mass)
of the gas velocity in our computations, instead using only
$\delta\hat{v}_r$.  The halo winds we use only conserve momentum
statistically, not per event, and so some dynamics may be lost there.
Furthermore, the amount of gas expelled from a halo and flowing over
it may depend on the halo mass itself. Lastly, even the no feedback
simulation has some regions with large relative velocities (see
Fig.~\ref{fig:slices}) and moreover has systematically heavier halos
which can complicate the interpretation of the mass dependence.  In
practice, we expect a combination of effects to be occurring.  We
discuss some of our efforts at disentangling them in Appendix
\ref{sec:disentangle}.

\section{Connection to Observables}
\subsection{The CDM Velocity}
Determining halo peculiar velocities from LSS observations is
notoriously tricky.  One way is to simply estimate it from linear
theory: $\vec{v}_L\propto -i\frac{\vec{k}}{k^2}\delta_m$.  Since
$\delta_m$ is not directly observable, the galaxy density field,
$\delta_g$, is used instead.  While the initial measurement of galaxy
coordinates can be quite precise, the can be difficult to use as it is
only a biased tracer of $\delta_m$.  Furthermore, the linear velocity
estimate is only accurate on sufficiently large scales.  Lastly, we
note that on sufficiently large scales both the gas and CDM have the
same linear velocity.  However, the inferred linear velocity field is
not directly sensitive to feedback, so we can extract the relative
velocity provided we have a direct measurement of the actual nonlinear
gas velocity.  An alternative approach is to measure the peculiar
velocities directly, which requires both a distance and a redshift and
is applicable only for low redshift galaxies \citep{bib:Hudson2012}.

One way to mitigate nonlinearities is through reconstruction
techniques which attempt to partially undo nonlinear gravitational
evolution.  These techniques work in Lagrangian space where the
relevant quantity is the displacement field rather than the density
field.  The standard reconstruction procedure
\citep{bib:Eisenstein2007} assumes the linear velocity field which we
wish to obtain, so that may not be useful.  However, there are a
number of nonlinear reconstruction methods
(e.g.~\citet{bib:Zhu2017,bib:Schmittfull2017,bib:Modi2018,bib:Hada2018,bib:Shi2018})
which use alternate methods to determine the displacement field.
Utilizing these methods increases the available information
\citep{bib:Pan2017} and has enabled better measurements of the baryon
acoustic oscillations \citep{bib:Wang2017}.  However, the success of
reconstruction in the presence of gas and feedback has not yet been
studied.

\subsection{The Gas Velocity}
In addition to traditional probes of the LSS such as the positions and
velocities of biased tracers, we may also consider the effects of
relative motions on the Sunyaev-Zel'dovich effects (SZEs).  The
thermal SZE is caused by the upscattering of photons by an ionized gas
in the Universe and is proportional to the cluster pressure.  These
regions also emit X-rays which can be used to determine the local gas
density \citep{bib:Adam2017}.  This is therefore directly related to
the sound speed,
$c_s^2 = \delta P/\delta\rho \sim \bar{P}/\bar{\rho}$, of the gas
around a halo.  The kinetic SZE arises when the ionized gas scattering
the photons has a net peculiar velocity.  Thus, its measurement tells
us the gas velocity around a halo.  Another possibility is to made use
of high-resolution X-ray spectroscopy to determine gas velocities
\citep{2018arXiv180706903G}. These measurements can also be used to
probe the gas velocity (inflows and outflows) our galaxies and
halos. The relative velocity changes significantly more than the gas
velocity with feedback and so its measurement may be more sensitive.
This then requires $v_c$, which could be obtained via linear density
reconstruction: $\vec{v}_c \sim \vec{v}_L$.  We note that the
simultaneous use of thermal and kinetic SZE alongside linear velocity
fields has already been performed \citep{bib:Schaan2016,
  bib:PlanckKSZ2016}.  Utilizing all three would require an excellent
understanding of the galaxy bias, however.Lastly, it is possible that
strong feedback such as the large relative velocities studied here
could bias estimations of the kinetic SZE \citep{bib:Park2018}.

\begin{figure}
  \includegraphics[width=\columnwidth]{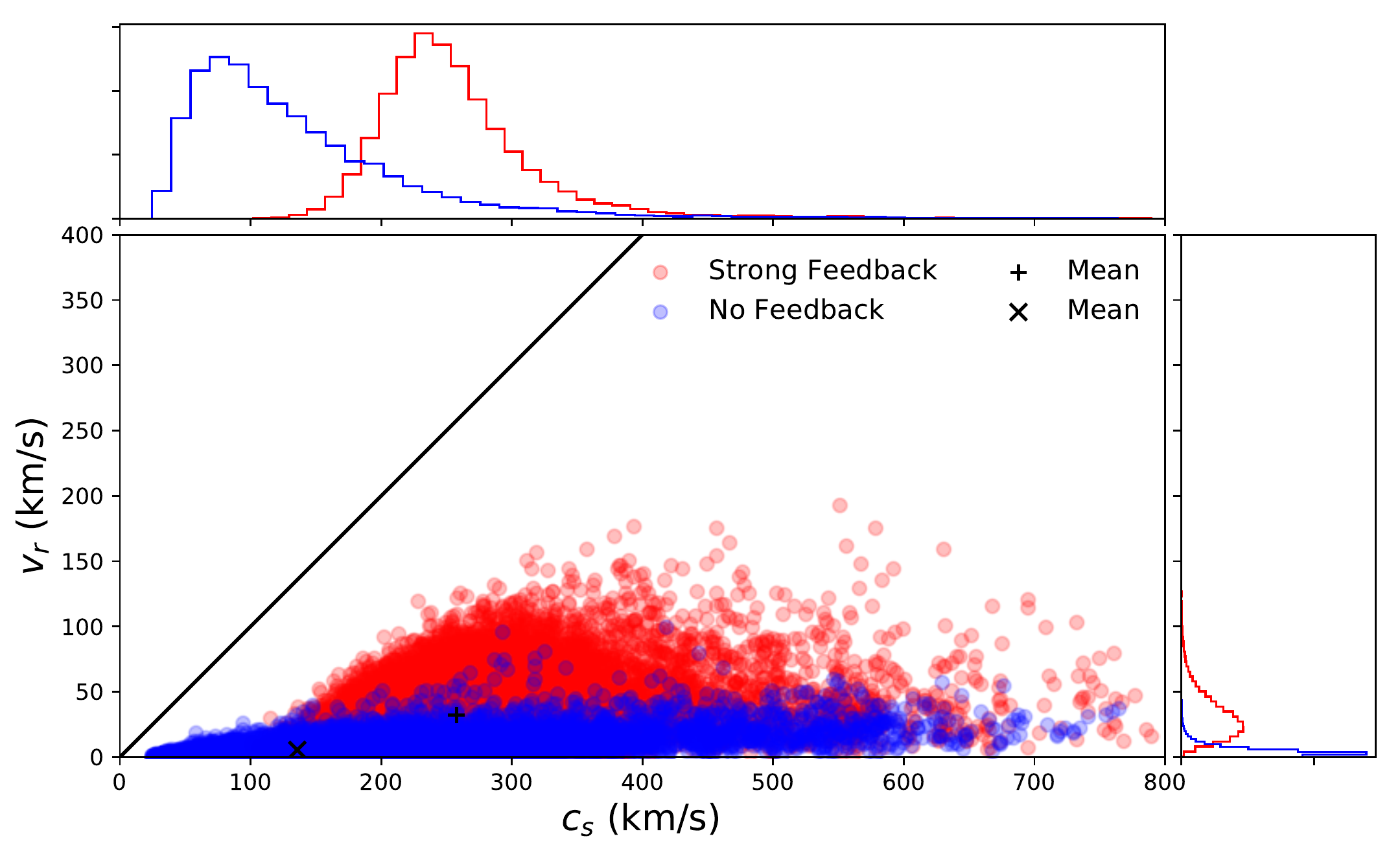}
  \caption{The relative velocity and sound speed computed in spherical
    shells of $5\ \mpch$ around each halo.  Strong feedback enhances
    both quantities significantly.  The top and right histograms
    project down to one variable.  The diagonal line is $v_r=c_s$ to
    guide the eye.}
  \label{fig:scatter}
\end{figure}

Nonetheless, it is plausible to obtain estimates of (or quantities
analogous to) $c_s$ and $v_r$ for individual halos.  Using our
results, we can make a theoretical prediction for what to expect.  We
estimate $c_s$, $\vec{v}_g$ and $\vec{v}_c$ by averaging all particles
in a sphere of $5$ \mpch{} as described in \S \ref{ssec:friction}.  We
then plot a phase diagram which we show in Fig.~\ref{fig:scatter}
where each point represents a halo.  Unsurprisingly, there is a clear
effect of strong feedback as there are significantly more halos with
large sound speeds and velocities.  This is also clearly shown in the
projected histograms where there is clear separation between the
simulations.  Finally, we see that $v_r \lesssim c_s$, in line with
Fig.~\ref{fig:vcz} and theoretical predictions
\citep{bib:Ostriker1999}.

\section{Conclusion}
\label{sec:conclusions}
We have studied some of the dynamical consequences associated with
feedback induced relative velocity between the IGM and CDM.  It is
worth keeping in mind a few things, however.  Firstly, we greatly
amplified the strength of supernova kinetic feedback, such that we
achieved winds of 1000s \kms.  This was done to mimic some of the more
energetic effects seen when AGN feedback is included (see, e.g.,
figures from the Illustris
\citep{bib:Vogelsberger2014b,bib:IllustrisTNG} or Eagle
\citep{bib:Schaye2015} simulations for striking examples) and to study
the dependence of our results on the feedback efficiency.  Secondly,
we have not spent much time discussing what happens if feedback is
low.  Here, the details of gas bias matter significantly, and will
likely be swamped by changes in the feedback model.  Dealing with
these considerations will require a more sophisticated feedback
implementation, but in principle, do not hinder our results.  A clear
next step is to see to what extent the effects presented here are
present in more realistic simulations.

It is also worth considering what other effects may mimic or
contaminate our results.  One (practically) guaranteed background is
from cosmic neutrinos, which will also have a relative velocity
\citet{bib:Zhu2014,bib:Inman2015,bib:Inman2017b} and dynamical
friction \cite{bib:Okoli2017}, although of course no SZE.  The numbers
we obtain may be compared to the prediction for neutrino dynamical
friction of \citet{bib:Okoli2017} who found values of order
$0.2\ \kms$ and $1.5\ \kpch$ (depending on halo and neutrino
properties) when averaged over $16$ \mpch.  Of course, the key
differences are that neutrinos are collisionless and therefore not
subject to hydrodynamic forces, but more importantly that the baryons
can contribute significantly more to the matter density since
$\Omega_g\gg \Omega_\nu$.  We therefore find that the IGM dipole
signal could be a significant contaminant to the neutrino one,
although it should not extend over as large of scales and the relative
velocities may be in different directions.  On the other hand, if
there is a warm component to the dark matter it could also contribute
and could be much more similar to gas.  On smaller scales (i.e.~$1$
\mpch{} or cluster sized scales), other effects come into play.  The
motions of galaxies through the halo can yield gravitational wakes
behind the galaxy in the direction of the relative motion
\citep{bib:Furlanetto2002}.  Furthermore, galaxies themselves will
feel Ram pressure \citep{bib:Gunn1972} due to the intracluster medium.
This should lead to a galaxy-gas dipolar effects which could be
enhanced if the ram pressure can strip gas from the galaxy
\citep{bib:Abadi1999}.

Finally, we re-emphasize that we have not considered more technical
challenges, such as the understanding of galaxy bias, the relationship
between galaxies and halos, and the computations of $v_r$ and $c_s$
from SZE measurements and galaxy surveys.  Certainly, more study is
required in this area.

\section*{Acknowledgements}
We thank David Spergel for valuable discussions about the IGM, Chia-Yu
Hu for discussions on how to compute the sound speed in hydrodynamic
simulations and Colin Hill, Nick Battaglia, and Pengjie Zhang for
useful discussions about the SZE. The numerical simulations have been
run in the \textit{rusty} cluster at the Flatiron Institute.  Parts of
the analysis were performed on the GPC supercomputer at the SciNet HPC
consortium \citep{bib:Loken2010}.  SciNet is funded by: the Canada
Foundation for Innovation under the auspices of Compute Canada; the
Government of Ontario; Ontario Research Excellence Fund - Research
Excellence; and the University of Toronto.  This work was supported in
part through the NYU IT High Performance Computing resources,
services, and staff expertise.  The work of FVN is supported by the
Simons Foundation. This research has made use of {\sc NumPy}
\citep{bib:Walt2011}, {\sc Matplotlib} \citep{bib:Hunter2007} and
NASA's Astrophysics Data System Bibliographic Services.

\bibliographystyle{mnras} \bibliography{thebib}

\clearpage
\appendix
\newpage
\section{Dynamical friction in the expanding universe}
\label{sec:exist}
An intriguing theoretical question is whether dynamical friction even
exists for a gaseous medium.  \citet{bib:Ostriker1999} computed the
force a point perturber feels and found that, in the steady state
limit, it is exactly zero for a gas in subsonic motion, but nonzero
for a collisionless species of particles.  Even if the steady state
limit is not achieved, a perturber that is turned on at some time will
feel a greatly reduced force when travelling subsonically.  This is
because the gas will try to produce a ``front-back symmetry,''
cancelling the drag force.  In this appendix, we will consider what
happens when this process occurs for fluids in an expanding Universe
filled with large scale structure.

The equation of motion for a fluid is given by:
\begin{align}
  \ddot{\delta}_g + w^2(a) \delta_g =
  \frac{3}{2}H_0^2\Omega_m \delta_m  a \nonumber
\end{align} 
where the frequency is given by $w(a)=akc_s(a)$ with $a$ being the
scalefactor and $k$ the comoving wavenumber.  In this equation,
over-dots refer to the Newtonian time, s, which is related to the
physical time, $t$, and conformal time, $\tau$, via $dt=ad\tau=a^2ds$.
For slowly varying sound speeds, we can employ the WKB approximation
and find the integral solution:
\begin{align}
  \label{eq:inteqn}
  \delta_g(\tau,k) = &G(s,s_i) \dot{\delta}_g(\tau_i) -
                       \dot{G}(s,s_i)\delta_g(\tau_i) \nonumber \\ &+\frac{3}{2}H_0^2\Omega_m \int_{\tau_i}^\tau d\tau'
                                                                     \delta_m(\tau',k)(s-s')\frac{G(s,s')}{(s-s')}
\end{align}
where the Green's function $G(s,s')$ is given by:
\begin{equation}
  \frac{G(s,s')}{(s-s')} = \frac{\sin\left(\int_{s'}^s ds'' w(s'')\right)}{
    (s-s') \sqrt{w(s)w(s')}} \nonumber
\end{equation}
and we have factored out the $s-s'$ piece such that in the limit of
$c_s\rightarrow 0$ that $G/(s-s')\rightarrow 1$.  The other two terms
in Eq.~\ref{eq:inteqn} arise due to the non-homogeneous boundary
conditions when $\tau_i$ is not $0$.  We will always take $\tau_i=0$
and so these terms are not present.  Eq.~\ref{eq:inteqn} is exact for
a gas with constant frequency ($c_s(a)=c_s/a$) where
$G(s,s')/(s-s')=j_0(k c_s (s-s'))$.  We will assume $c_s(a)=c_s/a$
throughout this Appendix.

Eq.~\ref{eq:inteqn} also applies to collisionless particles with a
given distribution function $f(v)$, provided initial conditions are
negligible and we use the appropriate Green's function:
\begin{equation}
  \frac{G(s-s')}{(s-s')} = \frac{\int dv v^2 f(v) j_0(k v (s-s'))}{\int dv v^2 f(v)}.\nonumber
\end{equation}
For the Maxwell-Boltzmann distribution (MBd),
$f(v)\propto \exp[-v^2/\sigma^2]$, and this integral can be done
analytically to obtain:
\begin{equation}
  \frac{G(s-s')}{(s-s')} = \exp\left[-\frac{\left(k \sigma (s-s')\right)^2}{4}\right]. \nonumber
\end{equation}
Because collisionless particles are dispersive \citep{bib:Inman2017a},
there is no value of $\sigma$ that matches $c_s$ at all scales.  The
closest definition to use is the small scale sound speed:
\begin{equation}
  c_s^2 = \frac{\int dv v^2 f(v)}{\int dv f(v)} \nonumber
\end{equation}
which is equal to $\sigma^2/2$ for the MBd.

To include the effects of a bulk flow we can use Moving Background
Perturbation theory \citep{bib:Tseliakhovich2010}.  A constant,
uniform bulk flow causes a comoving displacement,
$\vec{d}(\tau,\tau') =\int_{\tau'}^\tau \vec{v}_{\rm rel}(\tau'')
d\tau''$,
between the CDM and gas which becomes a phase in Fourier space
$\delta_c(k) \rightarrow \delta_c(k)\exp[i\vec{k}\cdot\vec{d}]$
\citep{bib:Inman2017b}.  To extract multipole moments, we can expand
the phase:
\begin{equation}
  \exp\left[i\vec{k}\cdot\vec{d}\right] = j_0(kd)+3i\mu j_1(kd)+...\nonumber
\end{equation}
The terms independent of $\mu$ constitute the monopole and terms
proportional to $\mu$ the dipole.  We can also expand the density
contrast into multipoles as well
$\delta = \delta_{0} + i\mu\delta_{1} + ...$.

To proceed further, we will work in matter domination and assume that
the gas or collisionless particles contribute negligibly to the matter
density: $\delta_m\rightarrow\delta_m\exp[i\mu k d]$.  In matter
domination we have $H_0=2/\tau_0$, $\Omega_m=1$,
$\delta_m(\tau,k) = a(\tau)\delta_m(1,k)$, $a(\tau)=(\tau/\tau_0)^2$
and $s-s'=-\tau_0^2(1/\tau - 1/\tau')$.  Finally we need to assume a
model for the induced displacement, $d$.  Since the sound speed is
decaying like $c_s\propto1/a$, the relative velocity should eventually
decay due to Hubble friction or else the bulk flows will become
infinitely supersonic.  We will therefore parameterize
$d = v_{\rm rel} (s-s') = M c_s (s-s')$ where $M$ is
the Mach number.

Using Eq.~\ref{eq:inteqn}, these
approximations, and changing variables to $x=\tau'/\tau$ yields the
following for the monopole and dipole:
\begin{align}
  \label{eq:delta0}
  \frac{ \delta_0(a,k) }{\delta_m(a,k)} &= 6\int_0^1
                                          x(1-x)V(\bar{k}(1-x)/x) j_0(M\bar{k}(1-x)/x)dx \\
  \label{eq:delta1}
  \frac{ \delta_1(a,k) }{\delta_m(a,k)} &= 6\int_0^1
                                          x(1-x)V(\bar{k}(1-x)/x)3j_1(M\bar{k}(1-x)/x)dx
\end{align}
where $\bar{k}=k c_s \tau_0/\sqrt{a(\tau)}$, and $V(x)=j_0(x)$ for gas
or $\exp\left[-x^2/2\right]$ for collisionless particles with a
MBd. Qualitatively, $V(x)\leq 1$ and so $\delta_0$ is never greater
than the CDM solution, $\delta_m$.  For the monopole in the subsonic
limit, $M \ll 1$, $j_0(M\bar{k}(1-x)/x)\to 1$ and we
expect the monopole to be unperturbed.  In the supersonic limit, $j_0$
will oscillate rapidly and we expect $\delta_0$ to become suppressed.
The dipole on the other hand is small (and proportional to
$M$) in the subsonic limit
$3j_1(M\bar{k}(1-x)/x)\to M\bar{k}(1-x)/x$.  It should also be
suppressed in the highly supersonic regime.

We can now see that the situation is very different in the
cosmological fluid equations used above compared to that of
\citet{bib:Ostriker1999}.  The sound speed experiences Hubble friction
and decays like $1/a$.  This implies that the Jean's wavenumber,
$k_J^2 = (3/2)H_0^2\Omega_m a/(a c_s(a))$ is increasing to smaller and
smaller scales.  This is reflected in our equations: taking
$a\to \infty$ implies $\bar{k}\to 0$ and $\delta_0\equiv\delta_m$
whereas $\delta_1\equiv 0$.  Thus, there will be no dipole distortions
(for linearized equations, anyways) in the infinitely far future, but
this is {\it not} due to the ``front-back'' symmetry found in
non-expanding fluids with a static perturber.  Instead, it is because
the hot fluids tend to the growing CDM solution.

\begin{figure}
  \includegraphics[width=\columnwidth]{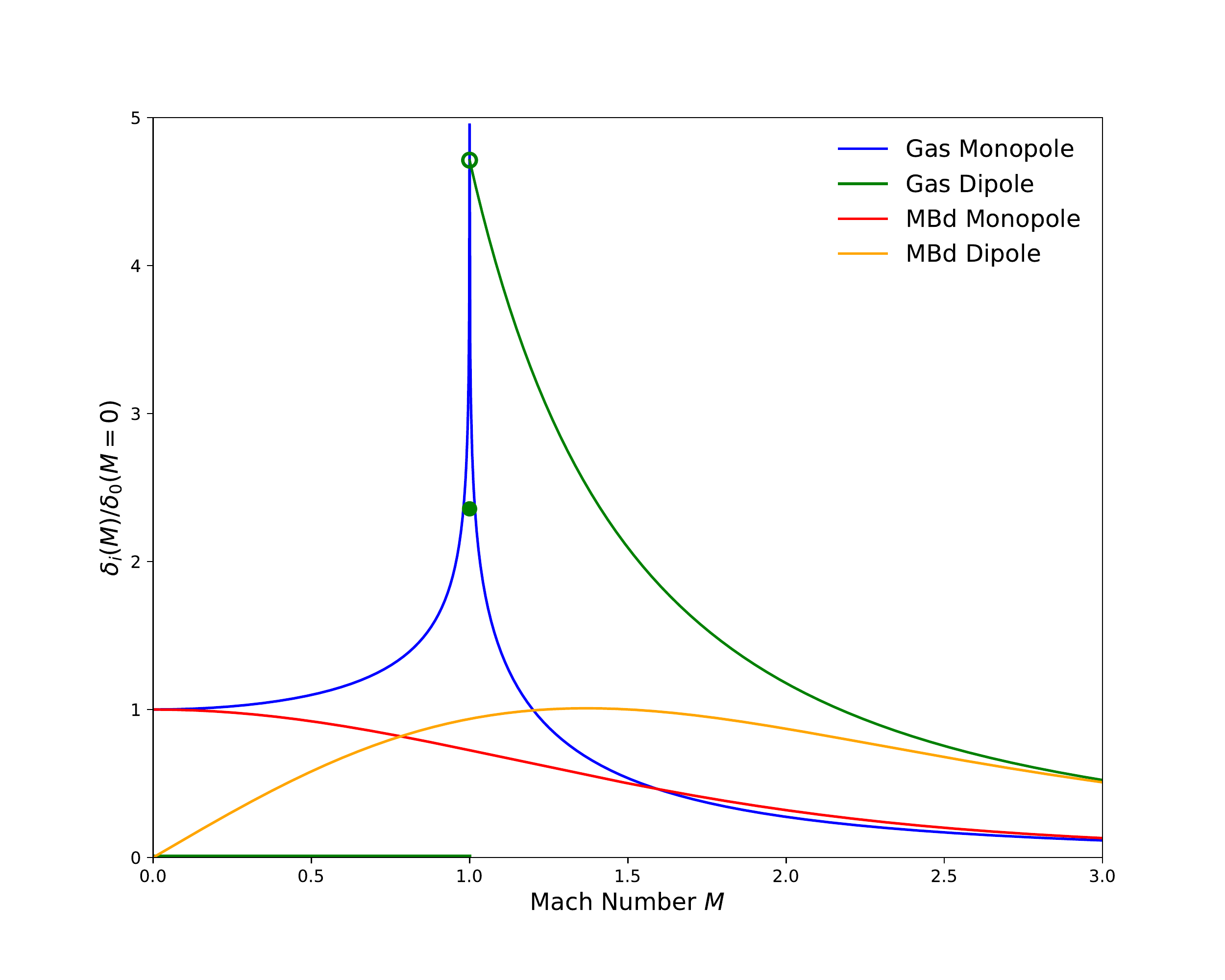}
  \caption{The instantaneous behaviour of the monopole and dipole
    relative to the zero relative velocity monopole as a function of
    Mach number.  Blue and red curves correspond to a collisional gas
    whereas red and orange are a collisionless gas with a
    Maxwell-Boltzmann velocity distribution.  Notably, the gas has no
    dipole for subsonic flows.}
  \label{fig:sstate}
\end{figure}

There is, however, a limit that much more closely matches the results
of \citet{bib:Ostriker1999}.  This limit is the ``instantaneous''
limit: $\ddot{\delta}_g \ll (kc_s)^2\delta_g$.  We can achieve this by
letting $\bar{k}\rightarrow\infty$.  To see the effects of this, it is
easiest to change variables in Eqs.~\ref{eq:delta0} and
\ref{eq:delta1} to $y=\bar{k}(1-x)/x$ in which case we obtain:
\begin{align}
  \label{eq:delta0lim}
  \frac{ \delta_0 }{\delta_m} &=\frac{6}{\bar{k}^2}
                                \int_0^\infty dy
                                \frac{y}{(1+y/\bar{k})^4}V(y)j_0(M
                                y) \\
  \label{eq:delta1lim}
  \frac{ \delta_1}{\delta_m} &=\frac{6}{\bar{k}^2}
                               \int_0^\infty dy
                               \frac{y}{(1+y/\bar{k})^4}V(y)3j_1(M
                               y).
\end{align}
For $M=0$, the limiting behaviours for Eq.~\ref{eq:delta0lim} is
$6/\bar{k}^2$, which is precisely the Jeans, or free streaming, scale
in matter domination: $\frac{k_{fs}}{k}=\frac{\sqrt{6}}{\bar{k}}$.
Eq.~\ref{eq:delta1lim} is exactly zero, as required.  Clearly, if we
take $\bar{k}\to\infty$ both $\delta_0$ and $\delta_1$ are zero for
non-trivial sound speeds.  However, we can understand their limiting
behaviour relative to the unperturbed ($M=0$) monopole,
i.e. $\lim_{\bar{k}\to\infty} \delta_i(M)/\delta_0(0)$ for $i=0,1$.
We show the result in Fig.~\ref{fig:sstate} for both gas and
collisionless MBd particles.  We find that the steady-state dipole is
exactly zero for subsonic collisional flows, but non-zero for
collisionless ones.  It is always non-zero for supersonic flows.  This
mimics the result seen in \citet{bib:Ostriker1999}: in the
instantaneous subsonic limit, the sound horizon of gas is effectively
infinite and there is no dipole.  Since there is no dipole in this
limit, there should also be no gravitational force to cause
backreaction in the CDM.  On the other hand, this is quite an artificial
limit and for finite $\bar{k}$ there will still be a dipole in the
subsonic case.

\section{Disentangling Halo Accelerators}
\label{sec:disentangle}
\begin{figure}
  \includegraphics[width=\columnwidth]{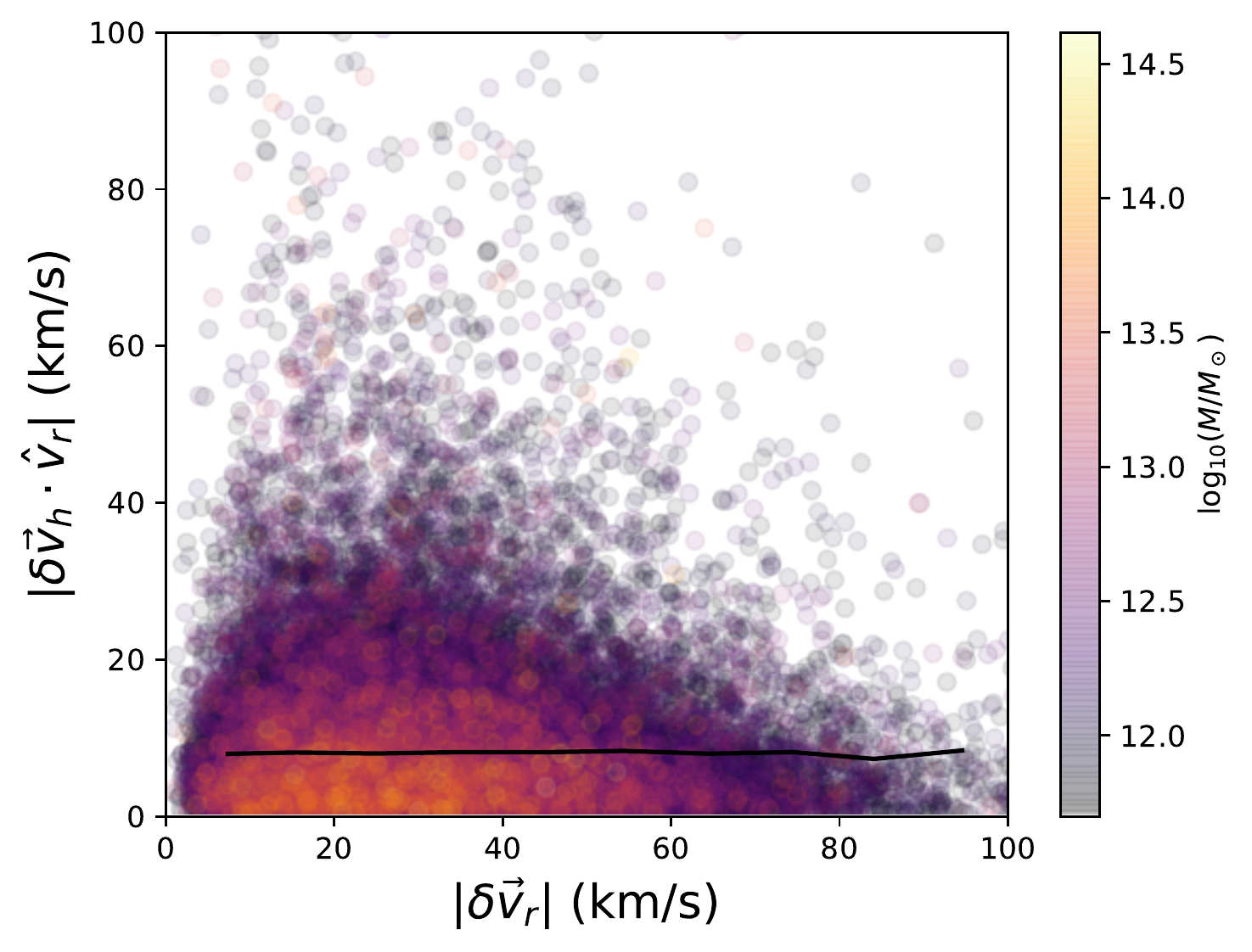}
  \caption{The differential acceleration of halos as a function of
    relative velocity.  The black line shows the mean value.}
  \label{fig:accv}
\end{figure}

The Ram pressure force scales like $\sim v^2$ \citep{bib:Gunn1972}
whereas dynamical friction is expected to fall off like $\sim 1/v^2$
\citep{bib:Ostriker1999}.  We therefore expect
$\delta \vec{v} \cdot \delta\hat{v}_r$ as a function of
$\delta \hat{v}_r$ to be different for the two effects.  We show this
result in Fig.~\ref{fig:accv}.  While by eye the shape looks
suggestive of dynamical friction, the mean value is very flat.

\begin{figure}
  \includegraphics[width=\columnwidth]{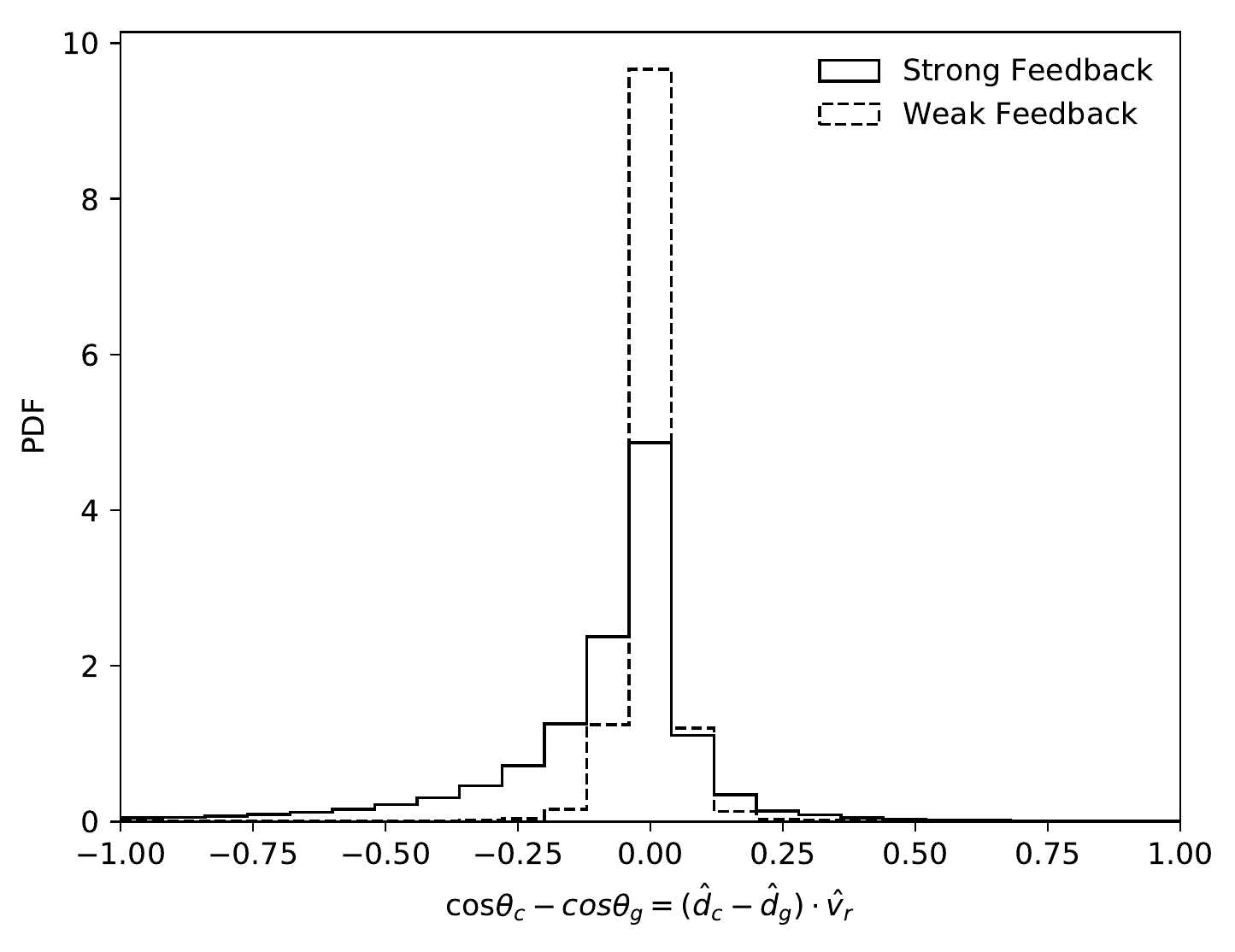}
  \caption{The alignment of the dipole and the relative velocity.}
  \label{fig:ctc}
\end{figure}

Alternatively, we could see gravitational in-fall of gas into a halo.
In this case we expect the halo to slightly move towards the gas, in
the opposite direction of the dynamical friction case.  To test this,
we compute the dipole moment of the density:
$\vec{d} \propto \sum (x_i-x_h)$ where $x_i$ are the coordinates of
particles within $5$ \mpch{} of a halo at $x_h$.  We then define
angles between the dipole and the relative velocity:
$\cos\theta_i=\hat{d}_i\cdot\hat{v}_r$.  We get two angles $i=c,g$ for
each simulation.  If dynamical friction is occurring, we expect the
gas dipole to be anti-aligned with the relative velocity, whereas it
will be aligned if it is gravitational in-fall.  In order to remove
contamination from bound gas in-falling with CDM, we compute
$\cos\theta_c-\cos\theta_g$ and show the result in Fig.~\ref{fig:ctc}.
The shift towards negative values is an indicator of in-fall.

\bsp	
\label{lastpage}
\end{document}